\documentclass[aps,prl,reprint,twocolumn,superscriptaddress,amsmath,amssymb,citeautoscript]{revtex4-1}
\usepackage{amsmath}
\usepackage{amssymb}
\usepackage{graphicx}
\usepackage{color}
\usepackage{bm}
\usepackage{epstopdf}
\usepackage{grffile}
\DeclareGraphicsExtensions{.eps}

\begin{document}

\title{Magnetic field induced criticality in superconducting two-leg ladders}
\author{Temo Vekua}
\affiliation{James Franck Institute, The University of Chicago, 60637 Chicago, Illinois} 

\begin{abstract}

  We study critical singularities in the d-wave-like superconducting phase of the hole-doped Hubbard model of repulsively interacting electrons, defined on a two-leg ladder, induced by a magnetic field applied parallel to the ladder plane. We argue that, provided the lowest energy spin excitations in doped ladders carry as well charge quantum numbers, the low temperature thermodynamic quantities, such as specific heat coefficient and magnetic susceptibility will show logarithmic singularities in quantum critical regime. This behavior is in drastic contrast with the magnetic field induced criticality in undoped Mott insulator ladders, which is governed by the zero scle-factor universality with its hallmark square root singularities.
\end{abstract}
\date{\today}

%%%%%%%%%%%%%%%%%%%%%%%%%%%%%%%%%%%%%
\maketitle

Ladder-like geometries are a minimal step from a purely one-dimensional (1d) structure towards two dimensions (2d), yet electron systems on ladders are amenable to being studied by powerful analytical and numerical methods available in 1d. In addition, these systems share some similarities with 2d behavior, e.g. upon doping the Hubbard model of repulsively interacting electrons on two-leg ladder develops pairing and dominant superconducting tendencies \cite{DagottoRiera,Hayward,Dagotto,Noack,Sigrist,Khveshchenko,BalentsFisher,Poilblanc,Roux,Troyer}. Superconductivity (that is the leading quasi-long-range order) in hole-doped ladders involves singlet pairing with an unconventional modified d-wave nature \cite{Sigrist,Khveshchenko,Noack,Poilblanc,Roux,Troyer}, reminiscent of behavior observed in high T$_c$ superconductors \cite{Tsui}. The fact itself that superconducting instability wins in purely repulsive system of electrons, without the phonon mediated attraction, may be suggesting the common origin of superconductivity in doped ladders and 2d high T$_c$ systems, with the magnetic fluctuations playing an important role \cite{Anderson}.

Two-leg ladder materials which show superconducting properties with doping, such as (La,Ca,Sr)$_{14}$Cu$_{24}$O$_{41}$, are synthesized, both as powder and as single crystals \cite{exp1,exp3}, fueling further interest in studying ladder systems.

Even though the spin gap is maintained with (not too large) doping ($\delta$) \cite{Noack}, it does not evolve continuously with $\delta$ \cite{Tsunetsugu,Lin,Roux}, effect which is beyond the mean-field approximation \cite{Sigrist}. Two scenarios have been suggested to explain this effect. In the first picture, in addition to conventional magnons, which evolve continuously with $\delta$, there are lower energy spin excitations: A hole-pair breaks up into two holes, each of which forms a bound state with up-spin electron on the same rung, producing spinon-holon quasi-particles  (carrying spin-1/2 and charge $|e|$) \cite{Tsunetsugu}, as illustrated in Fig. 1 (a). Another candidate for the lowest energy spin excitation in superconducting ladder is a bound state of magnon and hole-pair \cite{Lin,Roux}. In this scenario hole-pair does not break up, rather it gets dressed with magnon \cite{commentsecondscenario}. When a single hole-pair is present, it was argued that the singlet-triplet spin gap of the system, $\Delta_{tr}$, would experience a jump $\lim_{\delta \to 0}\Delta_{tr}(\delta) = (\sqrt{3}-1)\Delta_{tr}$ due to magnon-hole-pair bound state formation, and that this gap will further diminish with increasing $\delta$ \cite{Lin}. A gain in kinetic energy of holes in the local ferromagnetic environment of magnons was suggested as an intuitive mechanism for binding magnon to hole-pair \cite{Roux}. Numerical studies indicate that generically hole-pair is not tightly localized object on a rung, but rather it is spread over few rungs, and in particular, for isotropic hoppings, the maximum probability of the two holes configuration (participating in the bound state) is when they are located on the adjacent rungs and different legs \cite{Tsunetsugu,Siller}, as depicted on Fig, 1. A magnon-hole-pair bound state (carrying spin-1 and charge $2|e|$), corresponding to this case is sketched in Fig. 1(b) which is to be interpreted as: more depleted rungs are more polarized in the direction of the field.

\begin{figure}[h]
\includegraphics[width=7.0cm]{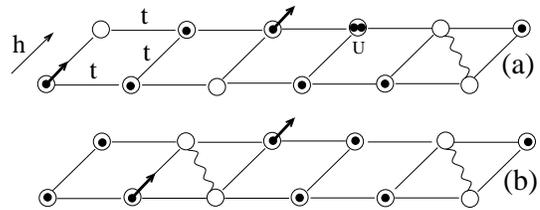}
\caption{ Cartoon of the lowest energy spin excitations in the doped Hubbard ladders with on-site interaction $U$ and hopping amplitudes $t$. Open circles denote lattice sites and filled circles represent electrons. In-plane magnetic field $h$ couples only to electron spins. On the last plaquette there is a hole-pair depicted and in (a) spinon-holons are depicted on first and third rungs while in (b) on second plaquette bound state of magnon and hole-pair is located.
}
\label{fig:patterns}
\end{figure}

Provided that the nature of the lowest energy spin excitations changes with doping, although the ground state continues to be spin singlet and a gap in spin excitation spectrum is maintained, a natural question arises what happens with the magnetic field induced quantum critical point with $\delta$, when magnetic field suppresses the singlet-triplet spin gap. The purpose of this work is to unveil universal singular properties of a magnetic field induced quantum critical point in a hole-doped ladder of repulsively interacting electrons with dominant superconducting instability.

Without doping, for half filling, and for strong on-site repulsion, the Hubbard model, at energies lower than the Mott gap, reduces to the Heisenberg spin-1/2 antiferromagnetic ladder. The lowest energy spin excitations in the absence of magnetic field are degenerate triplet of magnons, at wave-vector $(\pi,\pi)$, and are separated from the singlet ground state by an energy gap of $\Delta_{tr}$, which is roughly half of the Heisenberg exchange for the isotropic exchanges. Magnetic field splits the 3-fold degeneracy of the triplet bands linearly and at the critical value of magnetic field $h=\Delta_{tr}$ a phase transition is induced, and the ground state magnetization changes with magnetic field as $m_h=\frac{1}{\pi} \sqrt{h-\Delta_{tr}} \Theta(h-\Delta_{tr})$ \cite{Japaridze}, where $\Theta$ is the Heaviside step function. The phase transition induced by the critical magnetic field in undoped ladders shows the zero scale-factor universality \cite{Sachdev}, which in particular implies that both low temperature specific heat coefficient as well as magnetic susceptibility behave as $\sim T^{-\frac{1}{2}}$.

Perhaps the simplest non-trivial model of 1d electrons that shows gap in spin excitation spectrum is Hubbard model of attractively interacting electrons defined on a single chain, which is integrable and hence, one can obtain exact results, such as critical properties of magnetic field induced quantum phase transition. For attractive Hubbard model, using Bethe ansatz method, it was shown that away of half-filling critical ground state response of the system changes drastically as compared to the half-filled case \cite{Woynarovich86}. Similar results were obtained using 'non-linear' bosonization approach \cite{Vekua3}, where crucial role played by the spin-charge coupling, induced by the curvature of the Fermi surface, was unveiled. Using the Bethe ansatz basis of electron pairs and up-spin electrons, leading low temperature behavior of attractive Hubbard chain at magnetic field induced quantum critical point in dilute limit was obtained in \cite{Vekua1}.  

In the ladder problem, that we consider, pairing originates from repulsive inter-electron interactions and is hence truly many-body effect with Umklapp lattice processes playing an important role \cite{Khveshchenko}, as opposed to pairing in attractive Hubbard chain. In addition Hubbard ladders are not exactly solvable, and in weak-coupling bosonization description one has to start with four bosonic fields \cite{GNT} as compared to just spin and charge modes of the Hubbard chain. 

To address the nature of the magnetic field induced phase transition in doped superconducting ladders, we will need two assumptions for starting point: i) in the absence of magnetic field superconducting doped ladders have finite gap in spin excitations, while there is only one gapless mode: charge mode corresponding to the motion of hole pairs and ii) the lowest energy spin excitations are not conventional magnons, i.e. excitations involving only spin degree of freedom (which continuously evolve from undoped ladder case), but they involve charge degrees of freedom as well, similar to spinon-holon quasi-particles or magnon-hole-pair bound states. Consequently such composite objects (which carry both spin and charge quantum numbers) will start to populate the ground state once magnetic field overcomes the singlet-triple spin excitation gap of the doped ladders. 

Bosonization approach \cite{Lin}, supported by numerical simulations \cite{Tsunetsugu,Roux,Siller}, indicate that these ingredients are indeed met in lightly doped electron ladders, which maintain spin gap and show dominant superconducting correlations. It has been estimated numerically that spin gap survives at least until $\delta=0.25$ \cite{Noack}. In particular, it was suggested \cite{BalentsFisher,Tsunetsugu,Siller,Troyer} that lightly doped ladders at low energies realize Luther-Emery liquid \cite{LEL} and consequently hydrodynamic approach \cite{Larkin} was applied, producing an accurate low energy description of the system, in agreement with numerical simulations \cite{Tsunetsugu,Siller}. 

We adopt such hydrodynamic description of hole-pairs in doped spin ladders and as a new ingredient we add magnetic field via the Zeeman coupling, which is close to the critical value equal to singlet-triplet spin gap $h \simeq \Delta_{tr}(\delta)$. Magnetic field does not couple to hole-pairs, since hole-pairs do not carry spin. Rather, magnetic field acts as a chemical potential for the particles that represent low energy magnetic excitations that we will describe in effective continuous model by field $\psi$. As our aim is to describe leading low temperature behavior of the superconducting ladder in the vicinity of the critical magnetic field, for the effective model we may retain only those modes which are gapless (hole-pairs) and which will become gapless after magnetic field will exceed the critical value ($\psi$-particles). With these ingredients the effective theory governing low energy properties of the superconducting ladder in near-critical magnetic field looks,  
\begin{eqnarray}
\label{EFM}
{\cal H}&=&\frac{v_p}{2} \int \mathrm{d}x \left( \frac{ (n_p(x)-\bar n_p)^2}{K_p \pi}+ K_p\pi {(\partial_x \theta_p(x))^2} \right)\nonumber \\
&+& \int \mathrm{d}x \psi^{\dagger}(x) ( - \frac{\hbar^2 \partial_x^2 }{2M_{\psi}}-({h-h^0_{c}})s_{\psi} ) \psi(x) \\
&+&  \int \mathrm{d}x   \left[ g \psi^{\dagger}(x)  \psi(x) (n_p(x)-\bar n_p) + g_{\psi}  |\psi^{\dagger}(x)  \psi(x)|^2 \right] .\nonumber
\end{eqnarray}
We have fixed units $\hbar=1$ and introduced notation for spin value carried by the lowest energy spin excitation $s_{\psi}=1$ or $1/2$ depending on the case whether $\psi$ describes spin-$1$ magnon-hole-pairs or spin-$1/2$ spinon-holons. Hole-pairs are described by hydrodynamic variables, conjugate Gaussian fields ($n_p,\theta_p$) corresponding to density and phase fluctuations respectively, with canonical commutation relation $[n_p(x),\theta(y)]=i\delta(x-y)$. The Luttinger liquid constant $K_p$ as well as sound velocity $v_p$ can be estimated numerically \cite{Tsunetsugu,Siller}. In particular, Luttinger liquid parameter of pairs assumes universal value $K_p=1$\cite{Tsunetsugu,Troyer,Siller} for lightly doped case, meaning that hole pairs behave as hard-core bosons and $v_p\sim 1/\delta$. In the following we will set $K_p=1$ for simplicity.

The lowest energy spin excitations, which are either spinon-holon fermionic quasi-particles, or bosonic magnon-hole-pair particles, experience vacuum to finite density transition when magnetic field is swept across the critical value, and are described by either fermionic or bosonic field $\psi$. For the latter case, the last term in Eq. (\ref{EFM}) is important, with $g_{\psi}>0$ \cite{comment}. Since dilute limit is a strong-coupling limit in 1d, governed by Tonks gas fixed point, in the vicinity of critical value of magnetic field effectively $g_{\psi}\to \infty$ and hence we can treat magnon-hole-pairs as hard-core bosons.

The density-density interaction ($\sim g$) between the low energy modes in Eq. (\ref{EFM}) can shift the critical value of magnetic field $h_c^0\to \Delta_{tr}(\delta)$, however, it can not influence the nature of the underlying quantum critical point \cite{Balents} and in particular $g \to 0$ under renormalization. Eq. (\ref{EFM}) includes all terms up to quartic in fields allowed by symmetry of the microscopic problem. This is so in particular due to the $U(1)$ rotation symmetry in the spin space around the axes set by magnetic field and hence there can not be terms that contain odd number in $\psi$ field. Due to the very same reason, there must be equal number of $\psi$ and $\psi^{\dagger}$ operators in each term of effective Hamiltonian Eq. (\ref{EFM}), hence the so called 'pair hopping' processes between $\psi$ particles and hole-pairs, which can strongly influence the nature of phase transition \cite{Matveev} are excluded as well.

At critical magnetic field, $h=\Delta_{tr}(\delta)$, the low energy dispersion of $\psi$-particles is quadratic,
$E_{\psi}(k)= k^2/M_{\psi}$, invalidating their hydrodynamic description. From now on we will set $M_{\psi}=1/2$. The leading low energy dispersion of pairs though is linear in momentum, 
$E_p(k) \simeq v_p(|k|-k^p_F)\,\,   \mathrm{for}\,\, |k|\to k^p_F$, where the Fermi wave-vector of hole pairs is related to the linear density of pairs $\bar n_p$ by the standard 1d relation $k^p_F=\bar n_p \pi$.

For $h<\Delta_{tr}(\delta)$, number of $\psi$ particles in the ground state is zero. Once magnetic field exceeds $\Delta_{tr}(\delta)$, $\psi$-particles start to populate ground state by breaking up hole-pairs or dressing hole-pairs by magnons. Hence, even though magnetic field does not couple directly to hole-pairs, it alters quantum numbers of the ground state, in particular the mean ground state density of hole-pairs changes with magnetic field for $h>\Delta_{tr}(\delta)$ \cite{commentholedensiy} as, 
\begin{equation}
\label{const}
\bar n_{p} =\delta/2- {}_0\langle h| \psi^{\dagger} \psi |h\rangle_0 \,s_{\psi},
\end{equation}
where $|h\rangle_0$ is the ground state for a given value of magnetic field $h$, which is a global spin-singlet state and independent of field for $h<\Delta_{tr}(\delta)$. For $s_{\psi}=1$ one hole-pair is converted to one magnon-hole-pair, whereas for $s=1/2$ when one hole-pair breaks up two spinon-holon quasi-particles are produced. In both cases Eq. (\ref{const}) can be written in a unified manner,
\begin{equation}
\label{constraint}
\bar n_{p} =\delta/2-m_h.
\end{equation}

By minimizing the ground state energy, expectation value of the effective Hamiltonian Eq. (\ref{EFM}), with respect to $\bar n_{p}$, and expressing $\bar n_{p}$ with the help of Eq. (\ref{constraint}), we obtain for the ground state magnetization,
\begin{equation}
\label{gm}
m_h=\frac{\sqrt{4(h-\Delta_{tr}(\delta))/\pi^2 +v_p^2 }-v_p}{2}\Theta(h-\Delta_{tr}(\delta)).
\end{equation}
In particular, for $0<h-\Delta_{tr}(\delta)\ll v_p$ we obtain from Eq. (\ref{gm}), $m_h=\chi^S_c(h-\Delta_{tr}(\delta))+O((h-\Delta_{tr}(\delta))^2/v_p^2)$, where $\chi^S_c= 1/(\pi^2 v_p)\sim 1/\delta$ is the ground state critical spin susceptibility. Note, when $\delta=0$ the critical spin susceptibility diverges and one recovers from Eq. (\ref{gm}) the square root dependence of magnetization on the increment of magnetic field from the critical value. 

%In the following we will address leading low temperature properties of quantum critical point at $h=\Delta_{tr}(\delta)$. 
Even though at zero temperature for $h\le \Delta_{tr}(\delta) $ the ground state density of $\psi$-particles is zero, temperature fluctuations create spin excitations from different sources, however for very low temperatures only gapless excitations sources will be important, which involve hole-pairs. Hence, thermally induced magnetization for $h\le \Delta_{tr}(\delta)$ in the low temperature limit is,
\begin{equation}
\label{con}
m_{h}(T)=\delta/2-\bar n_p(T),
\end{equation}
where $\bar n_p(T)$ is the mean density of hole pairs at temperature $T$. As already mentioned, hole-pairs behave as hard-core bosons, at least in lightly doped case. Similarly $\psi$-particles, in case when they are bosonic, behave as hard-core bosons at the onset of magnetization, due to $g_{\psi}>0$ and the inherent strong-coupling nature of the critical point. In 1d, thermodynamic properties of hard-core bosons are identical to those of fermions, hence we can use Fermi-Dirac statistics for describing finite temperature distribution of both types of particles: hole-pairs as well as $\psi$-particles. We can rewrite Eq. (\ref{con}) with the help of the Lagrange multiplier chemical potential $\mu_h(T)$, which is a solution of the following equation,
\begin{equation}
\label{FTC}
\!\int_0^{\infty}\!\!\! \!\!\frac{dk/\pi}{ e^{\frac{ E_{\psi}(k)-h+\Delta_{tr}(\delta)-\mu_h(T) }{k_BT}}\!\!+\!1}+\!\!\int _0^{\infty}\!\!\!\!\!\frac{dk/\pi}{ e^{\frac{ E_p(k)-\mu_h(T) }{k_BT}}\!\! +\!1}= \frac{\delta }{2}.
\end{equation}
Solution for $\mu_{h}(T)$ from Eq. (\ref{FTC}) at $T\to 0$ is given in terms of the Lambert $\mathrm{W}$ function,
\begin{equation}
\label{LOG}
\frac{\mu_{h}(T)}{k_BT}=-\mathrm{W}\Big(\frac{ v_p e^{\frac{h-\Delta_{tr}(\delta)}{k_BT}} }{2\sqrt{k_BT/\pi}}\Big).
\end{equation}
Once $\mu_{h}(T)$ is known, within the grand canonical formalism, we obtain easily all interesting thermodynamic quantities of the system.

At critical field $h=\Delta_{tr}(\delta)$, the $\mu_{h}(T)$ picks up logarithmic dependence on temperature, as follows from Eq. (\ref{LOG}), dependence that carries to various thermodynamic quantities. In particular the critical specific heat coefficient behave as $\gamma_c(T)=C_c(T)/T\sim {\ln^2{T}}/v_p$ for $T\to 0$ and the leading low temperature dependence of critical magnetic susceptibility is, $\chi^S_c(T)-\chi^S_c\sim 1/\ln{T}$.

Hence, the critical properties of doped spin ladders in magnetic field that is equal to the singlet-triplet spin gap, differs drastically from undoped case. The crucial role is played by the occurrence of a new type of the lowest energy spin excitations with doping, which are composite in nature, carrying both spin and charge quantum numbers. When these objects start to populate the ground state, at the corresponding critical value of magnetic field, the spin susceptibility stays finite. The low temperature critical magnetic susceptibility as well as specific heat coefficients show logarithmic singularities, as opposed to the celebrated square root singularity of undoped ladders.

Magnetization curve and thermodynamic properties of the undoped ladders of organic compounds, whith a relatively small spin gap, have been measured in experiments \cite{exp4}. Similar studies have not been reported for the doped ladders, because of the difficulties of doping the organic materials. In doped non-organic ladder, in the superconducting phase, the spin gap of 80K has been reported \cite{exp3}, as compared to the spin gap of undoped system which is several hundreds of Kelvins. This gap is still too large to be suppressed by constant magnetic field available at present; However progress in achieving high magnetic fields and/or possibility (as suggested by theory) of further reducing spin gap of ladders by adjusting doping, make feasible experimental access of the character of magnetic field induced quantum critical point in superconducting ladders.

I am thankful to P. Wiegmann for helpful discussions and to G. Roux for relevant references. 
The work was supported in part by the NSF under the Grant NSF DMR-1206648.


\begin{thebibliography}{10}
\bibitem{commentsecondscenario} One can view magnon-hole-pair bound state sketched in Fig. 1(b) as a bound state of two spinon-holon quasi-particles depicted in Fig. 1(a). Thus difference between the two scenarios is whether the lowest energy spin excitation is carried by two spinon-holon scattering or bound state.
\bibitem{comment} From the site-resolved magnetization (and density) profile \cite{Roux} in lightly doped ladders it is clear that when magnetic field slightly exceed the critical value, magnon-hole-pairs repel each other, as they try to be distributed as far as possible from each other.
\bibitem{commentholedensiy} Note, even in the picture where hole-pairs get dressed by magnons, we still have formally to assume that hole-pair density drops for $h>h_c$ to avoid double counting, as $\psi$-particles include in them hole-pairs (bound to magnons).
\end{thebibliography}
\end{document}